\documentclass[11pt,a4paper, dvipsnames]{article}
\usepackage{jheppub}

\usepackage[utf8]{inputenc}
\usepackage{amsmath,setspace,amsfonts,latexsym}
\usepackage{amssymb}
\usepackage{color}
\usepackage{epsfig}
\usepackage{graphicx}
\usepackage{slashed}
\usepackage{comment}
\usepackage{nopageno}
\usepackage{cancel}
\usepackage{cite}
\usepackage{soul}
\usepackage{graphicx}
\usepackage{subcaption}
\usepackage{scalerel}

\usepackage[firstpageonly=true,hpos=0.63\paperwidth,vpos=0.82\paperwidth]{draftwatermark}

\SetWatermarkText{{\sf N}}
\SetWatermarkScale{52}
\SetWatermarkColor{gray!10!white}
\SetWatermarkAngle{0}

\usepackage{float}
\newcommand{\beq}{\begin{equation}\begin{aligned}{}}
\newcommand{\eeq}{\end{aligned}\end{equation}}
\newcommand{\beqa}[1]{\begin{equation}\begin{aligned}{#1}}
\newcommand{\eeqa}{\end{aligned}\end{equation}}

\usepackage[compat=1.0.0]{tikz-feynman}
\usetikzlibrary{decorations, decorations.text,backgrounds}

\newcommand{\betagw}{\beta_{\scaleto{\rm GW}{4pt}}}
\newcommand{\epsBZ}{\epsilon_{\scaleto{\rm BZ}{4pt}}}

\title{Supercooled Confinement}

\author{Prateek Agrawal,}
\author{Gaurang Ramakant Kane,}
\author{Vazha Loladze,}
\author{and Mario Reig}

\affiliation{Rudolf Peierls Centre for Theoretical Physics, 
University of Oxford, Parks Road, Oxford OX1 3PU, United Kingdom}
\emailAdd{prateek.agrawal@physics.ox.ac.uk}
\emailAdd{gaurang.kane@physics.ox.ac.uk}
\emailAdd{vazha.loladze@physics.ox.ac.uk}
\emailAdd{mario.reiglopez@physics.ox.ac.uk}

\abstract{
We study general properties of confinement phase transitions in the early universe. An observable gravitational wave signal from such transitions requires significant supercooling. However, in almost all understood examples of confining gauge theories the degree of supercooling is too small to give interesting gravitational wave signals. We review and highlight the evidence why supercooling is not generic in confining gauge theories. The exceptions are Randall-Sundrum models which define a strongly coupled gauge theory holographically by a 5D gravitational theory. We construct a simple illustrative model of a 4D gauge theory inspired by features of the Randall-Sundrum model. It is a large-$N$ gauge theory in the conformal window coupled to a weakly coupled scalar field which undergoes a supercooled phase transition that breaks the conformal symmetry and triggers confinement. We show that there are interesting features in the gravitational wave spectra that can carry the imprint of the confining gauge theory.
}

\begin{document}
\maketitle
\section{Introduction}
The detection of gravitational waves (GWs) has opened new avenues for exploring physics beyond the Standard Model (BSM)~\cite{Kosowsky:1992rz,Huber:2008hg}. The transparency of the universe to GWs allows us to probe very early epochs in cosmology, including temperatures above the weak scale. Many BSM scenarios may be probed through production of GW signals detectable by upcoming observatories \cite{Caprini:2024hue, Reardon:2023gzh}. Significant progress has been made in identifying various potential sources of these signals (see \cite{Caprini:2019egz}). 
In this paper, we focus on the prospect of GW production from a first-order confinement phase transition.

Multiple BSM scenarios with confinement phase transition have been discussed in the literature~\cite{Caprini:2019egz}. Strongly coupled theories with a confinement scale near or above the electroweak scale are compelling candidates for new physics, therefore it is crucially important to study the possible GW signal from the corresponding confinement transitions. One of the most well-motivated such theories are the composite Higgs models, where the compositeness in the Higgs sector can be motivated by the hierarchy problem \cite{Kaplan:1983fs,Georgi:1984af,Kaplan:1991dc,Contino:2003ve,Agashe:2004rs,Arkani-Hamed:2001nha,Chacko:2005pe}. 
A phase transition
near the electroweak scale is especially interesting because the peak frequency of the corresponding GWs naturally falls with the LISA sensitivity band \cite{Grojean:2006bp}.

In order to produce a sizeable GW signal, the phase transition should be first order. In addition, 
the signal is enhanced when the phase transition is strongly supercooled, indicating that the phase transition is strong and lasts long. In the specific case of confining phase transitions, the presence of strong coupling makes it challenging to extract the detailed physics of the phase transition. While there has been a lot of theoretical and numerical work to understand the confinement phase transition, often the focus of these studies is the thermal state near critical temperature. It is worth emphasizing that the physics of the metastable state and the degree of supercooling is much less well-understood, but drives the phenomenology of GW production.

Surprisingly, in well-understood corners of strongly-coupled 4D gauge theories which undergo confinement, supercooling seems to be elusive. We discuss various examples with evidence from lattice, theories with supersymmetry, and explicit holographic duals where we find that there is no appreciable supercooling. 
A well-studied example is pure Yang-Mills (YM) gauge theory in the large-$N$ limit, which undergoes a first-order confinement phase transition~\cite{Panero:2009tv}. However, this transition is not strongly supercooled and generates a negligible GW signal~\cite{GarciaGarcia:2015fol,Gouttenoire:2023roe}.

In the literature, confinement phase transitions in new physics models at the electroweak scale are modeled using gauge-gravity duality \cite{Megias:2018sxv,Agashe:2019lhy,Agashe:2020lfz,Bigazzi:2020phm,Bigazzi:2020avc}. The concrete gravitational dual theories are often taken to be the warped extra-dimensional Randall-Sundrum (RS) models~\cite{Randall:1999ee}. In the RS model the phase transition does exhibit strong supercooling. 
This motivates the question: what are the qualitative features in the RS model that allow for supercooling, and how can these be realized in the 4D gauge theory? Understanding this is crucial to establish if a measurable gravitational signal from a confining phase transition is generic.

In this paper, we construct and study an explicit field theoretical model in 4d that undergoes a first-order confinement phase transition and facilitates strong supercooling. 
We study a large-$N_c$ gauge theory with $N_f$ flavors of fermions near the lower end of the conformal window. The theory is coupled weakly to a scalar field which provides an external source of conformal symmetry breaking by giving mass to the fermions, driving the theory out of the conformal phase and triggering confinement. The scalar transition is first-order and undergoes strong supercooling, producing observable gravitational wave signals.

The structure of the paper is as follows. In section \ref{sec:appendixGW}, we start by reviewing some crucial aspects of GW signatures associated with cosmological first order phase transitions. Following this, we provide a brief discussion of the supercooled phase transition in a scenario with warped extra dimensions in section \ref{sec:RS}. In section \ref{sec:YM}, we review the literature on phase transitions in YM theory, arguing that in the minimal setup, it is likely the phase transition to be weakly first-order and thus incapable of generating a significant GW signal. Taking insights from section \ref{sec:RS}, we construct a simple field-theoretic model exhibiting a strongly supercooled first-order phase transition in section \ref{sec:model}. In section \ref{sec:causal_tails} we study the behaviour of GW signatures including the effects on causal tail from the phase transitions in our model. Finally, in section \ref{sec:discussion}, we explore the potential phenomenological implications of our study and summarize the results.

\section{Gravitational Waves from Cosmological Phase Transitions}\label{sec:appendixGW}

In this section we review the GW signals from cosmological phase transitions.
First-order phase transitions undergo by nucleation of bubbles. We define the nucleation temperature $T_n$ as the temperature at which the phase transition rate $\Gamma$ becomes significant:
\begin{align}
    \Gamma (T_n)
    &=
    H^4(T_n)\,.
    \label{eq:Tndef}
\end{align}
The degree of supercooling is defined in terms of the ratio of nucleation temperature $T_{n}$ to the critical temperature $T_{c}$ (the temperature below which the phase transition is allowed),
\begin{align}
    \epsilon_n
    &=
    1-\frac{T_n}{T_c}\,.
    \label{eq:epsilon_n}
\end{align}
A phase transition is strongly supercooled for $\epsilon_n \sim 1$. Following nucleation, the bubbles expand and collide filling up the whole universe with the new phase. The bubble collisions generate gravitational waves. There are three main sources of GW production identified in the literature: the collision of bubble walls \cite{Kosowsky:1991ua, Kosowsky:1992vn}, turbulence in the plasma \cite{Kosowsky:2001xp,Caprini:2006jb,Caprini:2009yp}, and sound waves \cite{Hogan:1986dsh,Hindmarsh:2013xza,Hindmarsh:2015qta,Hindmarsh:2016lnk}.

The phase transition as well as the resulting production of GWs is a very complicated process and thoroughly understanding them requires numerical simulations. An important quantity parameterizing the GW signal is the inverse duration of the phase transition, $\betagw$. In a rough approximation, one can parametrize the transition rate below temperature $T_n$ as~\cite{Turner:1992tz}, 
\begin{align}\label{eq:beta_rate}
 \Gamma 
 &= 
 H^4(T_n) e^{\betagw (t-t_n)}   
 \,.
\end{align}
In terms of the Euclidean action, $\betagw$ is defined as
\begin{align}
   \betagw
   &=
   \frac{1}{\Gamma}\frac{d\Gamma}{dt}\approx H_n\left(T\frac{d}{dT}S_{\scaleto{\rm E}{4pt}}-4\right)\Bigg\vert_{T=T_n}\,.
   \label{eq:beta}
\end{align}
However, for confinement phase transitions, the equations~\eqref{eq:beta_rate} and \eqref{eq:beta} are only accurate in the regime where the phase transition is not strongly supercooled. Hence, in the case of strongly supercooled phase transition one should use a better definition of $\betagw$ which involves the inverse characteristic size of bubbles at the time of collision~\cite{Hindmarsh:2019phv} 
\begin{align}
    \betagw = (8 \pi )^{1/3} \frac{v_w}{R^*}
    \,,\,\,\, \text{with\ } 
    R^* =\left(n_b^{*}\right)^{-1/3}\,.
    \label{eqn:beta_proper_def}
\end{align}
Here $n_b^*$ is the average number density of bubbles at the time of collision, defined as $I(t^*)=1$ (see equation~\eqref{eq:I_def}), and $v_w$ is the bubble wall velocity. The number density of bubbles can be determined by solving the equation
\begin{align}
    \frac{dn_b}{dt} = \Gamma(t) e^{-I(t)} \,.
\end{align}

Along with $\betagw$, another crucial parameter that sets up the overall strength of the GW signal is $\alpha$. This parameter roughly measures the amount of energy in the phase transition relative to the total energy of the universe and is defined as
\begin{equation}
\alpha= \frac{1}{\rho_{\scaleto{R}{4pt}}(T_p)}\left(\Delta V- \frac{T}{4}\Delta\frac{dV}{dT}\right)\Bigg\vert_{T=T_{p}}\approx \dfrac{\Delta V}{\rho_{\scaleto{R}{4pt}}(T_{p})}  \,,
\label{eqn:alpha}
\end{equation}
where $\rho_{\scaleto{R}{4pt}}$ is the radiation energy density, $\Delta V$ is the energy difference between the two vacua and $T_p$ is percolation temperature (defined below). The final approximation in the equation above is valid in case of a supercooled phase transition. Strong GWs require large $\alpha$, raising the issue of possible vacuum energy domination that may affect the transition such that it may never complete. 

The usual criterion for percolation  
is that the probability of finding a point in space to be in the false vacuum at temperature $T < T_c$ is given by $e^{-I(T)}\approx0.71$ or
\begin{align}\label{eq:I_def}
I(T_p)&=
\int_{T_p}^{T_c}
\frac{dT'}{T'}
\frac{\Gamma(T')}{H(T')}
\frac{4\pi}{3}(a(T') r(T_{p},T'))^3
\approx 0.34\,,
\end{align}
where
\begin{align}
r(T,T')
&=
\int_T^{T'}
\frac{d\widetilde{T}}{\widetilde{T}}
\frac{v_w}{H(\widetilde{T}) a(\widetilde{T})}\,.
\end{align}
In addition, a large value of $\alpha$ together with a large duration of the phase transition ($\betagw/H_{\star}\sim 10$) implies that the bubbles could be inflated away before colliding. One can ensure that the transition completes by checking that at the time of percolation $t_p$, found using equation~\eqref{eq:I_def}, the volume in false vacuum bubbles decreases~\cite{Turner:1992tz},
\begin{equation}
    \frac{1}{V_{\rm false}}\frac{dV_{\rm false}}{dt}\bigg|_{t=t_p}=3H(t_p)-\dfrac{dI}{dt}\bigg|_{t=t_p}<0~~.
\end{equation}

Finally, another important parameter that determines the behavior and strength of the phase transition is the bubble wall velocity $v_w$, introduced above in equation~\eqref{eqn:beta_proper_def}. In weakly coupled scenarios, a strong PT ($\alpha\gtrsim 1$) ensures relativistic wall velocities, $v_w\rightarrow 1$ \cite{Espinosa:2010hh}. However, the determination of the velocity in strongly coupled scenarios may involve additional complications.

We now review the GW spectra arising from the sources mentioned above. These spectra are empirical fits to numerical simulations. The GW spectrum from the bubble wall collisions can be calculated using the envelope approximation \cite{Kosowsky:1991ua,Kosowsky:1992vn}. In the envelope approximation, the energy fraction carried by GWs scales with $\betagw$ and bubble wall velocity $v_w$ as \cite{Caprini:2015zlo, Hindmarsh:2013xza, Hindmarsh:2015qta, Hindmarsh:2016lnk, Hindmarsh:2017gnf, Hindmarsh:2019phv, Giblin:2014qia, Lewicki:2021pgr, Cutting:2018tjt} 
\begin{align}
 h^2\dfrac{d\Omega_{\rm env}}{d\ln f}
 &= 
 1.67\times 10^{-9}
 \dfrac{0.11 v_{w}^{3}}{0.42+v_{w}^{2}}\left(\dfrac{100}{g_{*}}\right)^{\frac{1}{3}}\left(\dfrac{100}{\betagw/H_{\star}}\right)^{2}
 \left(\dfrac{\kappa \alpha}{1+\alpha}\right)^{2}
 S_{\rm env}\left({f/f_{\rm env}}\right)\,,
 \nonumber \\
 S_{\rm env}(x)
 &=
 \dfrac{3.8 x^{2.8}}{1+2.8x^{3.8}}\,.
\end{align}
Here, $f_{\rm env}$ is the peak frequency of the spectral function and is characterised by the inverse duration $\betagw$. The parameter $h = H_0/(100 {\rm\ km / s / Mpc})$ is the Hubble parameter today. The peak frequency is defined at the time of GW production and redshifts as radiation from the phase transition epoch to today. $H_{\star}$ is the Hubble during the GW production, $g_{*}$ is the number of relativistic degrees of freedom in the plasma, $\kappa$ is the fraction of latent heat that is deposited in the thin shell close to the phase transition.

An estimate of the GW signal generated from sound waves is~\cite{Caprini:2015zlo} 
\begin{align}
  h^2\dfrac{d\Omega_{\rm sw}}{d\ln f}
  =&2.65\times 10^{-8}v_{w}\left(\dfrac{100}{g_{*}}\right)^{\frac{1}{3}}\left(\dfrac{100}{\betagw/H_{\star}}\right)\left(\dfrac{\kappa_{v} \alpha}{1+\alpha}\right)^{2} 
  S_{\rm sw} (f/f_{\rm sw})\,,
  \nonumber \\
  S_{\rm sw} (x)
  &=
  x^3\left(\dfrac{7}{4+3x^{2}}\right)^{7/2}\,,
  \label{eqn:gw_fraction_sound_waves}
\end{align}
where the peak frequency can be estimated to be $f_{\rm sw}\simeq(2/\sqrt{3})(\betagw/v_{w})$. This estimate assumes that the bubble wall velocity $v_{w}$ is more that about $10\%$ away from the speed of sound or light. Further, $\kappa_{v}$ is the efficiency factor of the latent heat that transforms into fluid bulk motion and can be calculated from $\alpha$.

The GW spectrum from turbulence is estimated to be 
\begin{align}
    h^2\dfrac{d\Omega_{\rm turb}}{d\ln f}
    &=
    \small 3.35\times 10^{-6}v_{w}\left(\dfrac{100}{\betagw/H_{\star}}\right)\left(\dfrac{\kappa_{\rm turb}\alpha}{1+\alpha}\right)^{\frac{3}{2}}\left(\dfrac{100}{g_{*}}\right)^{\frac{1}{3}}
    S_{\rm turb}(f)\,,
    \nonumber \\
    S_{\rm turb}(f)
    &=
    \dfrac{(f/f_{\rm turb})^{3}}{[1+(f/f_{\rm turb})]^{\frac{11}{3}}+(1+8\pi f/h_{*})}~~.
\end{align}
Similar to the sound wave the peak frequency of the signal $f_{\rm turb}$ is estimated to take the form $f_{\rm turb}\simeq 1.75(\betagw/v_{w})$.  Further, $\kappa_{\rm turb}$ is the efficiency factor that parametrizes the fraction of latent heat of phase transition that goes into turbulence and $h_{*}$ is the Hubble rate redshifted to today calculated assuming that the universe was dominated by radiation immediately after the phase transition. For further details on $\kappa_{\rm turb}, \kappa_{v}$, we refer the readers to \cite{Espinosa:2010hh,Caprini:2015zlo}. 
From the relations above, it is clear that a large GW signal requires large $\alpha$, relativistic $v_w$, and a small $\betagw/H_{\star}$. The inverse duration $\betagw$ is inversely correlated with the degree of supercooling $\epsilon_{n}$. 
We will see this explicitly in section~\ref{sec:YM} in the case of YM theory and in section~\ref{sec:numerical_analysis} in the case of our model. For a loud transition where $(\betagw/H_{\star}) \simeq 100$, one  requires $\epsilon_n \sim 1$. The GW contribution is dominated by the sound waves and turbulence contributions over the collision of the walls \cite{Hindmarsh:2015qta}.

\section{Supercooling in the Randall-Sundrum Model} \label{sec:RS}

In the last section, we argued that strong GW signals can be produced if the phase transition is supercooled (i.e.~$\epsilon_n\lesssim 1$). We are interested in phase transitions related to confinement and their possible GW signals. In this section, we review a well-known holographic example where supercooling is obtained, namely the Randall-Sundrum (RS) model\cite{Randall:1999ee}. The RS model  also serves as a compelling solution to the hierarchy problem. The model consists of an $\text{AdS}_5$ spacetime compactified on a $S^{1}/Z_2$ orbifold, with IR and UV three-branes situated at the orbifold fixed points. In this setup, the Planck-weak hierarchy is achieved through warping of the fifth dimension. The RS metric is
\begin{align}
ds^2
&= k^2 \rho^2 dt^2 
- \frac{d\rho^2}{k^2 \rho^2}
-\rho^2 k^2 dx_i^2
\label{eq:RS_metric_1}
\end{align}
where $k$ is the AdS$_5$ curvature and the UV and IR branes are at fixed coordinates $\rho_{\rm uv}$ and $\rho_{\rm ir}$.

In a thermal bath the temperature in the 5D theory varies along the warped direction. A useful definition of the temperature of the theory is the temperature at the UV brane.
The RS branes are stabilized using a Goldberger-Wise field \cite{Goldberger:1999uk}. This mechanism stabilizes the IR brane in the RS solution in equation~\eqref{eq:RS_metric_1} at low temperatures. At high temperatures, the solution falls behind the horizon of the AdS-Schwarzschild (AdS-S) black hole~\cite{Creminelli:2001th}. The metric for the AdS-S solution is
\begin{align}
    ds^2
    &=
    f(\rho) dt^2 - \frac{d\rho^2}{f(\rho)}
    - \rho^2 k^2 dx_i^2,
    \qquad
    f(\rho) = f^2\left(\rho^2 -\frac{\rho_h^2}{\rho^2}
    \right)\,,
\end{align}
where $\rho_h = \pi T_{h} /k^2$ denotes the position of the AdS-S horizon.  While at high temperatures the gravitational partition function is dominated by the AdS-S solution, below some critical temperature the free energy of the RS solution becomes smaller than that of the AdS-S solution, causing the transition from the AdS-S to RS phase.

Through the gauge/gravity duality, the AdS-S solution corresponds to a conformal field theory coupled to 4D gravity at finite temperature in the deconfined phase, and the RS solution describes the theory in the confined phase\cite{Rattazzi:2000hs}.  The phase transition described above is interpreted as a confinement transition in the dual gauge theory~\cite{Creminelli:2001th}.

The confinement transition can be studied in the limit of $y_{\rm uv}\to -\infty$ which corresponds to a decoupling of 4D gravity \cite{Creminelli:2001th}. If the effects of the Goldberger-Wise mechanism on the phase transition are ignored, the AdS-S solution is found to dominate the partition function at any finite temperature. This can be understood as follows: in the absence of the stabilization mechanism, the RS as well as AdS-S solutions describe a conformal field theory on $R^3\times S^1$. In this case, the theory at any finite temperature is always in the deconfined phase. It is worth drawing a contrast with the case of $\mathcal{N}=4$ super YM~\cite{Witten:1998zw}, where the CFT is placed on a sphere. The sphere provides a scale, so that there can be a phase transition when the temperature scale crosses the scale of the sphere radius.

This phenomenon is also reflected in the potential for the radion in the RS model without Goldberger-Wise stabilization. The potential for the radion is exactly zero when both the UV and IR brane tensions are tuned. The IR brane can be placed anywhere in the AdS bulk, corresponding to spontaneous breaking of conformal invariance. The spontaneously broken conformality also fixes the form of the finite temperature radion potential to be a runaway potential with no stable minimum. 
The Goldberger-Wise mechanism induces an effective potential for the radion field through a breaking of conformal invariance. When this effect is included in the phase transition calculation, an AdS-S to RS transition becomes possible.  

The transition involves the black hole horizon receding and emergence of the IR brane. In RS models, the part of the tunneling action responsible for the change of spacetime topology and black hole geometry is neglected (and argued as being subdominant for some choice of parameters) and the action is parametrized in terms of the transition of the IR brane from the far IR to its stabilized value. The phase transition is thus calculated as a scalar field transition using the radion potential. 
The radion Lagrangian including the effective potential due to the Goldberger-Wise field can be written as:
\begin{align}
\mathcal{L}_{\rm radion}
&=
\frac{N^2}{16\pi^2}
\left((\partial \varphi)^2 
-\lambda (g(\varphi)) \varphi^4
\right)\,.
\end{align}
The explicit factor of $N$ corresponds to the identification of the radion as a composite glueball state in the gauge theory description. The conformal symmetry breaking effects of the Goldberger-Wise mechanism are encoded in the $\varphi$ dependent coefficient of the $\varphi^4$ potential of the radion. 

The free energy in the confined phase is essentially the potential of the radion at its minimum, $\varphi=f$,
\begin{align}
    F_{\rm confined}
    &=
    V(f) = V_0 - \frac{N^2}{16\pi^2} \lambda_f f^4 \,.
\end{align}
The deconfined phase free energy is given by the thermal distribution of the $N^2$ degrees of freedom in the gauge theory
\begin{align}
    F_{\rm deconfined}
    &=
    C - \frac{\pi^2}{96} N^2 T^4 \,,
\end{align}
where $C$ is a constant to match the free energy across the phase transition. The critical temperature can be estimated as,
\begin{align}
    T_c
    &\simeq \lambda_f^{1/4} f\,.
\end{align}

To calculate the phase transition rate, one needs to calculate the bounce action. One of the major challenges while calculating the bounce action in this scenario is that one needs to know the full gravitational solution with the bubble. In this case, as argued in \cite{Creminelli:2001th}, the bounce action is dominated by the non-gravitational contribution. The bounce actions calculated in \cite{Creminelli:2001th, Agrawal:2021alq} are:
\begin{align}
    S_{4}\sim \dfrac{N^{2}}{16\pi^{2}\lambda_{f}}~,~~~~~\dfrac{S_{3}}{T}\sim \dfrac{N^{2}}{8}\left(\dfrac{1}{\lambda_{f}}\right)^{\frac{3}{4}}\dfrac{(T_{c}/T)}{\left(1-(T/T_{c})^{4}\right)^{2}}
     \label{eqn:bounce_action_rattazzi}~~.
\end{align}
In the case of RS, the quartic $\lambda_{f}$ depends on the details of the stabilization, and is parametrically smaller than 1 for the simplest cases, but more sophisticated stabilization models and phase transitions in them have been studied, see for example\cite{Agashe:2020lfz,Mishra:2023kiu,Mishra:2024ehr}.
We see from expressions in equation~\eqref{eqn:bounce_action_rattazzi} that for large $N$ and small $\lambda_f$ the bounce actions are large and hence the transition can be supercooled or may not complete. Such long periods of supercooling may have important implications for the cosmology of dark matter models as studied in \cite{Baratella:2018pxi,Baldes:2020kam,Baldes:2021aph}.

It is worth emphasizing the critical features in the RS setup that allow for large supercooling. Despite being strongly coupled, the theory allows for a scale separation due to the proximity of an interacting fixed point. The phase transition to a deconfined phase is only possible due to a breaking of conformality. However, in cases where the deformation is relevant and it is a big deformation that drives the confinement (e.g.~in QCD the theory is very far from scale invariant at the confinement scale), then strong running effects lead to instability (related to instabilities seen in references \cite{Buchel:2021yay} and \cite{Mishra:2024ehr}). When the deformation is irrelevant, or remains small, then confinement is ``mostly spontaneous", stabilized by the small explicit breaking~\cite{CPR, Coradeschi:2013gda,Bellazzini:2013fga, Agrawal:2016ubh}. In this case the deconfined phase is near-conformal and can undergo sizable supercooling without any instability. This is precisely the cases for the most-studied RS models.

The RS model provides an interesting benchmark for a strongly coupled gauge theory with supercooled confining transition. The RS model dynamics are understood in the holographic setting on the gravitational side, or with the effective radion (glueball) Lagrangian in the confined phase. The explicit 4D gauge theory that leads to \emph{spontaneous confinement} seems difficult to construct and to our knowledge such a construction does not exist in the literature. This motivates looking for a 4D gauge theory model that can reproduce these features. In section~\ref{sec:model} we construct a model that reproduces the qualitative features of the RS phase transition and can produce loud GWs.

\section{The Confinement Phase Transition in the Yang-Mills Theory}\label{sec:YM}

In this section we highlight the issues that explain why 
achieving large supercooling within the minimal framework of a confinement transition in YM theory is challenging, if not impossible. 

\subsection{Large-\texorpdfstring{$N$}{N} confinement from the Lattice}
The deconfinement/confinement phase transition in pure YM theory with a vanishing vacuum angle has been extensively studied on the lattice. It is well-established that the phase transition in pure $SU(N_{c})$ gauge theory with $N_{c}\geq 3$ is of the first order \cite{Panero:2009tv}, and therefore proceeds via bubble nucleation. To estimate the phase transition rate, we can apply the thin-wall approximation and derive the $O(3)$ symmetric action for a bubble of radius $R$
\begin{eqnarray}
    S_3=4\pi R^2 \sigma-\frac{4\pi}{3}R^3\Delta f\,,
\end{eqnarray}
where $\sigma$ is bubble wall tension and $\Delta f=f_d-f_c$, with $f_d$ -- free energy density in deconfined phase and $f_c$ -- free energy density in confined phase. The action is extremal for the critical radius, $R_c=2\sigma/\Delta f$. By evaluating the action at the critical radius we get
\begin{eqnarray}
    S_3=\frac{16\pi}{3}\frac{\sigma^3}{\Delta f^2} \,.
\end{eqnarray}
We can obtain $\Delta f$ up to first order in ($1-T/T_c$) from the latent heat of the phase transition, $L$,
\begin{align}
    \Delta f(T)&=
    \Delta f(T_c)+\Delta f'(T_c)(T-T_c) 
    = 
    L\frac{T_c-T}{T_c} \,.
\end{align}
The quantities $L$ and $\sigma$ can be extracted from lattice simulations~\cite{Lucini:2005vg}: 
\begin{align}
    L
    &=
    \left(0.766(40)-\frac{0.34(1.60)}{N_{c}^2}\right)^4 N_{c}^2T_c^4 \,,
    \\
    \sigma
    &=
    \left(0.0138(3) - \frac{0.104(3)}{N_{c}^2}\right) N_{c}^2 T_c^3
    \,.
\end{align}
This gives us the $O(3)$ symmetric bounce action in large-$N$ limit,
\begin{eqnarray}
\frac{S_3}{T}\approx 3.7\times10^{-4}\frac{N_{c}^2}{\epsilon(T)^2}\left(1-\frac{7.5}{N_{c}^2}\right)^3 \,,
\label{eqn:o_3_ym}
\end{eqnarray}
with $\epsilon(T)=\frac{T_c-T}{T_c}$. The phase transition rate per unit volume is
\begin{eqnarray}
    \Gamma(T)=Ae^{-S_3/T}\,.
\end{eqnarray}
In the expression, $A$ is the fluctuation determinant which is $O(T_c^4)$. The nucleation temperature $T_n$ is defined in equation~\eqref{eq:Tndef}, which allows us to calculate the degree of supercooling
\begin{eqnarray}
    \epsilon_n
\approx 1.6\times10^{-3}N_{c}\left(1-\frac{7.5}{N_{c}^2}\right)^{3/2}
\left(
1+\frac{1}{140}\log{\left(\frac{A}{T_c^4}\right)}
    \right)^{-1/2} \,.
    \label{eq:nucleationtemp}
\end{eqnarray}
Note that expression is obtained in the limit $\epsilon_n\ll 1$ and $T_c \sim 1{\rm\ TeV}$. We see from this equation that even for $N_{c}\sim 10$ the supercooling is extremely small $\epsilon_n\sim 2\times 10^{-2}$. 

To analyze the GW signal, it is essential to carefully study the dynamics of the phase transition. 
Since the degree of supercooling is small, generically, the latent heat released during bubble expansion heats the universe back up to $T_c$. At this point the pressures of confined and deconfined phases become equal and the bubble expands with a velocity just enough to maintain a constant temperature, with the energy removed with Hubble expansion compensated by the released latent heat. This restricts the bubble velocity to be very small and severely suppresses GW signal~\cite{GarciaGarcia:2015fol,Gouttenoire:2023roe}. 

On the other hand, if we imagine that the latent heat is removed by another sector with a very large heat capacity
we can neglect the impact of the released energy on the thermal evolution of the universe and the bubble dynamics. However, in this case the existence of the additional sector again severely decreases the GW signal because in this scenario only a fraction of the degrees of freedom in the thermal bath participate in the phase transition.
Using equation~\eqref{eq:beta} one can additionally estimate that in this case $\betagw/H_{\star}=200/\epsilon_n\sim 2\times 10^{4}$, suppressing the signal further.
Therefore in either case, we do not expect a strong GW signal from the confinement phase transition.

We might wonder if increasing $N_{c}$ further could increase supercooling. For instance, if $N_{c}\gtrsim 10^2$ we find from equation~\eqref{eq:nucleationtemp} that $\epsilon_n\gtrsim 0.2$.  However, this equation was derived under the condition $\epsilon(T) \ll 1$. To understand for what values of $\epsilon(T)$ is the calculation reliable, note that the derivation of equation~\eqref{eq:nucleationtemp} assumes a domain wall tension $\sigma$ measured at $T_c$. If we expand the domain wall tension as $\sigma(T) = \sigma(T_c) + \sigma'(T_c) T_c \epsilon(T) + \mathcal{O}(\epsilon^2)$, the natural size of correction is $\sigma'(T_c) \sim \mathcal{O}(N_{c}^2 T_c^2)$. Given the parameterically small value of $\sigma(T_c)$, the calculation remains reliable only for $\epsilon(T) \lesssim 10^{-2}$. Therefore, based on only existing lattice data we cannot reliably answer the question if increasing $N_{c}$ beyond $N_{c}\sim 10$ makes the transition strongly supercooled. In the next subsection, we study well-known examples using AdS/CFT correspondence to get some intuition about the amount of the supercooling for $N\gg 10$.

We have discussed the case of phase transition in pure YM theory at $\theta=0$. One can further investigate to whether one gets large supercooling in presence of $\theta$ angle or by adding fermions in the theory. 
However, to the best of our knowledge, obtaining supercooling by adding fermions or by having a non-zero $\theta$ angle is still an open question. Keeping this in mind, we assume a vanishing vacuum $\theta$ angle and leave the $\theta\neq 0$ case for future investigation. 
\subsection{\texorpdfstring{$\mathcal{N}$}{N}=4 SYM using AdS/CFT}
In section~\ref{sec:RS} we reviewed the case of supercooling in the RS model where the dual gauge theory description in not known. In this section, we take a look at the holographic cases where the dual gauge theory is precisely known, although we emphasize that these models do not have the same phenomenology as RS. We start by reviewing the simplest, most well-known example where confinement can be studied through the gauge/gravity duality: the \(\mathcal{N}=4\) supersymmetric YM theory on \(S^3 \times S^1\) with circumferences \(\beta_{S^{3}}\) and \(\beta_{S^{1}}\), respectively. As the only relevant quantity in the theory is $\beta_{S^{1}}/\beta_{S^{3}}$ we can always set $\beta_{S^{3}}=1$. 

The thermodynamics of this theory is well understood by analyzing its gravity dual in \(AdS_5\) with a boundary \(S^3 \times S^1\) \cite{Witten:1998zw}.
The gravitational theory in Euclidean space has two classical solutions. The thermal AdS solution represents the confined phase of the CFT,
\begin{align}
    ds^2
    &=
    \left(
    \frac{r^2}{L_{\rm AdS}^2}+1\right)dt^2
    +\frac{dr^2}{\frac{r^2}{L_{\rm AdS}^2}+1}
    +r^2d\Omega^2 \,,
\end{align}
where $L_{\rm AdS}$ is the AdS curvature radius and $d\Omega^2$ is the metric on the unit-radius $S^3$.
The black hole solution corresponds to the deconfined phase,
\begin{align}
    ds^2
    &=
    \left(\frac{r^2}{L_{\rm AdS}^2}+1-\frac{r_h^2}{r^2}\right)dt^2+\frac{dr^2}{\frac{r^2}{L_{\rm AdS}^2}+1-\frac{r_h^2}{r^2}}+r^2d\Omega^2 \,,
\end{align} 
where $r_h$ is the event horizon. The black hole solution is regular only if
\begin{eqnarray}
    r_h=\frac{\pi^2L_{\rm AdS}^2}{2\beta_{S^{1}}}\left(1\pm \sqrt{1-\frac{2\beta_{S^{1}}^2}{\pi^2L_{\rm AdS}^2}}\right) \,.
    \label{eq:schwarzschildrad}
\end{eqnarray}
Thus, there are two possible values of \(r_h\) for each \(\beta_{S^{1}}\). 
At high temperatures, the black hole solution with the larger radius (corresponding to the ``$+$'' sign in equation~\eqref{eq:schwarzschildrad}) dominates the partition function of the canonical ensemble, while at low temperatures, the thermal AdS solution is the dominant one. The smaller-radius black hole corresponding to the ``$-$" sign has negative specific heat and never dominates. The first-order phase transition between the black hole and thermal AdS, known as the Hawking-Page transition~\cite{Hawking:1982dh}, occurs below the critical temperature
\begin{align}
T_c &= \frac{d-1}{2\pi L_{\rm AdS}} = \frac{3}{2\pi L_{\rm AdS}}\,,
\end{align}
in AdS$_{d+1}$ ($d=4$ for us). In the dual CFT, this corresponds to a confinement / deconfinement transition. 

We see from equation~\eqref{eq:schwarzschildrad} that there is a maximum value of \(\beta_{S^{1}}\) for which black hole solutions exists, \(\beta_{S^{1},\rm max}=\pi L_{\rm AdS}/\sqrt{2}\). On the CFT side, this implies that the deconfined phase only exists above the minimum temperature
\begin{equation}
T_{\rm min}
= \left(\beta_{S^{1},\rm max}\right)^{-1}
=\frac{\sqrt{d(d-2)}}{2\pi L_{\rm AdS}}
=\frac{\sqrt{2}}{\pi L_{\rm AdS}}~~.   
\end{equation}
This implies that the maximum supercooling during the confinement transition can be: 
\begin{align}
  \epsilon(T_{\rm min})=1-\dfrac{T_{\rm min}}{T_c} 
  = 1 -\frac{\sqrt{d(d-2)}}{d-1}\approx 0.057\,.
\end{align} 
Therefore, even if the first-order transition is suppressed near the critical temperature, the transition is inevitable at lower temperatures, limiting the degree of supercooling. Intriguingly, the smallness of $\epsilon(T_{\rm min})$ is set by a geometrical factor,  $1/d^2$.

\subsection{The Klebanov-Strassler duality cascade}
The $\mathcal{N}=4$ SYM example shows that the degree of supercooling in that case is bounded by an instability, parametrically similar to the small supercooling in YM on the lattice. However, as noted, in $\mathcal{N}=4$ SYM the confinement transition is driven by putting the CFT on a sphere so that the low-energy theory has ``kinematic confinement" due to the constraint of putting the gauge theory on a compact space. This may be different from confinement on $R^3$ driven by running couplings. 

The Klebanov-Strassler duality cascade 
\cite{Klebanov:2000hb} is a holographic example with $\mathcal{N}=1$ supersymmetry with 
a running coupling where confinement is driven by the dynamics. It is a controlled example close to the RS model, where we can study the gauge theory on Minkowski space. The confinement transition exists even in infinite volume limit, making it more similar to pure YM theory. In the gravity dual of the theory, the confined phase is described by the black brane that becomes unstable below some $T_{\rm min}$ such that $\epsilon(T_{\min})\sim 0.1$  \cite{Buchel:2009bh}.

The transition can also be studied in the framework of improved holography that similarly to the theories considered above exhibits the existence of the minimal temperature below which a deconfined phase can not exist. In this case we also have $\epsilon(T_{\rm min})\sim 0.05$ \cite{Morgante:2022zvc}. 
In all the  examples discussed above there is a minimum temperature $T_{\rm min}$ below which the deconfined phase does not exist. Hence, there is a maximum limit to supercooling that can be achieved in these cases characterized by $T_{\rm min}$ after which the transition has to happen and is governed by instabilities.  Thus we might expect that pure $SU(N)$ with very large $N$ does not have large supercooling either. Of course, this discussion above is not a formal proof but taking into account the physics described here this seems  plausible. 

It is worth mentioning that phase transition triggered by instabilities could proceed in a qualitatively different way. As shown in \cite{Bea:2021zol}, the transition could occur via the formation and merging of phase domains with distinctive phenomenology in the form of a qualitatively different GW spectrum.

\section{A Simple Field Theory Model}\label{sec:model}
We have seen that achieving a supercooled confinement transition in a gauge theory is a challenging task. In this section, we introduce an illustrative model which facilitates a first-order confinement transition with substantial supercooling. We use lessons from the AdS-S to RS transition discussed in section~\ref{sec:RS} as a guidance. We consider a CFT and couple it to an additional weakly coupled sector that breaks conformal invariance in a controlled way. In our concrete model it corresponds to a YM theory with quarks in the fundamental representation. The breaking of conformal invariance is achieved by a scalar field that undergoes a first-order phase transition and gets a non-vanishing vacuum expectation value (vev). We will show quantitatively how the phase transition in the scalar sector controls the amount of supercooling in the confinement phase transition. The concrete model that we consider is chosen to illustrate the principles of model building to achieve a supercool confinement phase transition.

\subsection{The Model}
We study a $SU(N_{c})$ gauge theory with $N_{c}\gg1$ and $N_{f}$ fermions in the fundamental representation. It is believed that there exists a $N_{\rm con}$ such that for $N_{\rm con}<N_{f}<11N_c/2$ the theory runs to a non-trivial IR fixed point. This is the so-called conformal window. At the upper end of the window, the  Banks-Zaks (BZ) fixed point \cite{Banks:1981nn} emerges for $N_f=11N_{c}/2-N_{c}\epsBZ$ with $\epsBZ\ll 1$ at weak coupling. The lower end of the conformal window is more difficult to pin down, since the fixed point will be at strong coupling. It is expected that $N_{\rm con}\sim \mathcal{O}(N_{c})$ because we know $N_{\rm con} <11N_{c}/2$ and the theory does not have any small parameter other than $1/N_{c}$ itself (for the ratio $N_{\rm con}/N_{c}$ to survive in the large-$N$ limit, it should be $\sim 1$). This intuition works very well in the context of $\mathcal{N}=1$ supersymmetric QCD where the value can be predicted using superconformal algebra as $N^{\scaleto{\rm SQCD}{5pt}}_{\rm con}=3N_{c}/2$. This comes from the consistency conditions of the superconformal theory, and it is not a rigorous proof of the existence of fixed points in the window but strong evidence of it~\cite{Seiberg:1994pq}.

The existence of the fixed point in non-supersymmetric YM theory is studied using different methods which put the lower end of the conformal window at $N_{\rm con}=(3-4) N_{c}$ \cite{Appelquist:1996dq,Gies:2005as, Jarvinen:2011qe, Ryttov:2017lkz, DiPietro:2020jne}. The IR behavior of the $SU(N_{c})$ gauge theory is therefore given by:
\begin{enumerate}
    \item For $N_f\geq 11N_{c}/2$ the theory is IR free.
    \item For $N_{\rm con}<N_f<11N_{c}/2$ the theory runs to a non-trivial IR fixed point where it can be described as CFT.
    \item For $N_f\leq N_{\rm con}$ the theory confines. 
\end{enumerate}

Here we present a simple modification of $SU(N_{c})$ gauge theory in the conformal window which has a supercooled confinement phase transition\footnote{Note that we could also consider the case with $N_f\geq 11N_{c}/2$. The physics of this regime is very similar to what we study here with the additional complication that the gauge coupling runs.}. We introduce an extra real scalar field $\phi$ which couples to fermions with the same (or comparable) Yukawa coupling $y$. The goal is to generate a temperature-dependent potential $V(\phi,T)$ for $\phi$ such that at high temperature it has a unique global minimum at $\phi=0$ while below a critical temperature, $T<T_c$, the minimum at $\phi=0$ becomes a local minimum separated by a barrier from a global minimum $\langle \phi \rangle$ which gives a mass $m_\psi = y\langle \phi \rangle$ to the fermions.  This drives the theory away from the conformal fixed point, and the theory will have a confinement scale close to $m_\psi$. The bubbles start being nucleated at an appreciable rate at temperature $T_n$.  If the scalar transition temperature $T_n$ is smaller than the YM confinement temperature $T_{\rm con}$, $m_\psi \gtrsim T_{\rm con} > T_{n}$, the theory undergoes confinement simultaneously with the scalar phase transition. 
This scenario is realized by the Lagrangian
\begin{eqnarray}
    &&\mathcal{L}=\mathcal{L}_{YM}+\frac{1}{2}(\partial_\mu\phi)^2-V_{\rm tree}(\phi)-y \phi\sum_{n=1}^{N_F} \bar{Q}^{n}Q^{n} \,, \nonumber \\
    &&V_{\rm tree}(\phi)=\frac{\lambda}{4}\phi^4+\frac{\phi^6}{6\Lambda^2}\,.
    \label{eq:treelevelpotential}
\end{eqnarray}
Here $\mathcal{L}_{YM}$ is the Lagrangian of gauge theory with $N_{c}$ color and $N_{f}$ quarks $Q^i$, and $\Lambda$ is a scale of new physics responsible for the non-renormalizable interaction. Running of the quartic is important and will be discussed together with renormalization scheme below in section~\ref{sec:numerical_analysis}.

We choose the potential to not have a mass term for simplicity. At the phase transition temperatures the scalar has a thermal mass and an addition of a bare mass term will not qualitatively change the results. The dimension-6 operator is present to lift the potential at large $\phi$ values. It parametrizes the effects of UV physics at the scale $\Lambda$ and more general choices will not change our conclusions.
The potential in equation~\eqref{eq:treelevelpotential} has a $\mathbb{Z}_{2}$ symmetry that can lead to domain wall formation. It is easy to imagine additional operators that break this $\mathbb{Z}_{2}$ symmetry without affecting any of the phase transition dynamics.

We need to make sure that the coupling of $\phi$ with quarks does not destabilize the fixed point in the YM theory.
In the subsection to follow, we study this issue and show that the fixed point can be maintained in the parameter space of our interests at least at one-loop order.

\subsection{The Fixed Point}
The existence of a fixed point for the gauge theory is crucial for this mechanism. In this section we study how the scalar field $\phi$ affects the fixed point.
We start by studying the fixed point in the perturbative regime and the scalar field expanded near $\langle\phi\rangle=0$. At $y=0$ the two loop beta function of the gauge coupling is given by \cite{Jones:1974mm, Caswell:1974gg}:
\begin{eqnarray}
    &&\beta(g)=-g\left(\beta_0\frac{g^2}{16\pi^2}+\beta_1\frac{g^4}{(16\pi^2)^2}\right) \,,\nonumber \\
    &&\beta_0=\frac{11}{3}N_{c}-\frac{2}{3}N_{f} \,, \nonumber \\
    &&\beta_1=\frac{34}{3}N_{c}^2-\frac{10}{3}N_{c}N_{f}-2C_2(F)N_{f} \,,
\end{eqnarray}
where $C_2(F)=\frac{N_{c}^2-1}{2N_{c}}$ is the quadratic Casimir of the fundamental representation. The BZ fixed point is obtained at
\begin{eqnarray}
    \frac{g_{\star}^2N_{c}}{16\pi^2}=\frac{\epsBZ}{3\beta_1(N_{f}=11N_{c}/2)/N_{c}^2}\,,
\end{eqnarray}
with $\epsBZ=11-2N_f/N_{c}\ll 1$. 

Now let us consider a small deviation from the fixed point $\Delta g=g-g_{\star}\ll g_{\star}$. The Renormalization Group (RG) equation to leading order in $\Delta g$ has the form: 
\begin{eqnarray}
    \mu\frac{\partial}{\partial\mu}\Delta g=2\beta_0\frac{g_{\star}^2}{16\pi^2}\Delta g \,.
\end{eqnarray}
It is clear from the equation that along RG flow to IR $\Delta g\to 0$ and the fixed point is stable. 
Turning on the Yukawa coupling modifies the RG  equation as:
\begin{eqnarray}
        \mu\frac{\partial}{\partial\mu}\Delta g=2\beta_0\frac{g_{\star}^2}{16\pi^2}\Delta g+a\frac{N_{f}g_{\star}^3}{(16\pi^2)^2}y^2 \,,
        \label{eq:runinggwithy}
\end{eqnarray}
where the last term comes from two loop diagrams involving a $\phi$-loop and has been estimated up to an order one number $a\sim \mathcal{O}(1)$. Thus the fixed point can be maintained up to $\Delta g/g_{\star}\sim y^2$ corrections. Note that for the perturbative treatment to be valid, we need $y N\lesssim 4\pi$ (where $N^{2}=N_{c}N_{f})$. Hence the model is calculable if the coupling $y$ is of the order $y\sim 1/N$ meaning that we have a fixed point at $g=g_{\star}$ up to $1/N^2$ corrections.

We consider now the Yukawa and quartic couplings of the scalar. The one-loop RG equation for $y$ and $\lambda$ are given by: 
\begin{eqnarray}
    &&\mu\frac{\partial}{\partial\mu}y^2N^2=\frac{1}{8\pi^2}y^2N^2\left(\left(3+2N^2\right)y^2-6C_2(F)g_{\star}^2\right)\,, \nonumber \\
    &&\mu\frac{\partial}{\partial\mu}\lambda=\frac{9}{8\pi^2}\lambda^2+\frac{y^2 N^2}{2 \pi^2}\lambda-\frac{N^2}{2\pi^2}y^4 \,,
\end{eqnarray}
where we assume that $g=g_{\star}$. The equations show that in the limit $y\sim 1/N$, to leading order in $1/N$ expansion, the quartic has fixed points at
\begin{align}
\lambda_\pm
&=
\frac{2y^2N^2}{9}\left(-1\pm\sqrt{1+9/N^2}\right),
\end{align}
where $\lambda_-$ is the UV fixed point and $\lambda_+$ is the IR fixed point. Similar to the Higgs quartic in the SM, the quartic coupling runs negative in the UV.
At leading order in $1/N$,
\begin{align}
\lambda_{-}
&=
-\frac{4y^2N^2}{9}
\, , \quad
\lambda_{+} = y^2\,.
\label{eq:quarticfixed}
\end{align}
Therefore, the theory has three fixed points:
\begin{enumerate}
    \item $g=g_{\star}$, $N^2y^2_{\star}=3N_{c}g^2_{\star}/2$, and $\lambda_-=-4y_{\star}^2N^2/9$. 
    
    \item $g=g_{\star}$, $N^2y^2_{\star}=3N_{c}g^2_{\star}/2$, and $\lambda_+=y_{\star}^2$ . 
    \item $g=g_{\star}$, $\lambda_0=4y_{\star}^2N^2/9= 0$. 
\end{enumerate}
In the next section, we will study the confinement phase transition in the theory.

\subsection{The Phase Transition}

In this section, we investigate the phase transition in the model by exploring the parameter space where the theory exhibits the two relevant phases: (1) $\langle\phi\rangle=0$, where the YM theory is at its IR fixed point (and hence deconfined), and (2) $\langle\phi\rangle \neq 0$, where the fermions become massive and the fixed point is destabilized. If the quark mass in the true vacuum is much smaller than the phase transition temperature then confinement happens at much lower temperature $T\sim\mathcal{O}(m_\psi)$, which is well separated from the scalar transition. Instead, we will be interested in a scenario where, in the true vacuum phase, the quarks are massive enough $m_\psi\gtrsim T_c$ to entirely decouple from the gauge theory dynamics. In this case, the scalar phase transition can trigger confinement.

\subsubsection{The Thermal Potential}

In this section we study the phase transition at finite temperature. First, we describe the potential that we consider for the phase transition. Following that, we perform numerical analysis using the {\tt CosmoTransitions} package \cite{Wainwright:2011kj}. We introduced the tree-level potential in equation~\eqref{eq:treelevelpotential}. Following \cite{Curtin:2016urg}, we include the one-loop RG improved Coleman-Weinberg \cite{Coleman:1973jx} potential. Along with this, we consider one-loop thermal corrections with resummed thermal mass in the truncated full dressing scheme. 

We will work in the RG improved scheme.
The potential can be split up into the tree-level potential $V_{\rm tree}(\phi)$, the temperature-dependent Coleman-Weinberg potential $V_{\scaleto{\rm CW}{4pt}}(\phi,T )$, and the thermal effective potential $V_{\scaleto{T>0}{5pt}}(\phi, T)$. The full potential is written as:
\begin{align}
V(\phi,T)=V_{\rm tree}(\phi)+V_{\scaleto{\rm CW}{4pt}}(\phi,T)+V_{\scaleto{T>0}{5pt}}(\phi,T)\,,
\end{align}
where, the tree-level potential has the form:
\begin{align}
    V_{\rm tree}(\phi)=\frac{\lambda(\mu)}{4}\phi^4+\frac{\phi^6}{6\Lambda^2}\,,
\end{align}
and the Coleman-Weinberg potential is given by:
\begin{equation}
 V_{\scaleto{\rm CW}{4pt}}(\phi,T)\small =\frac{\left(m^2(\phi)+\Pi(T,\phi)\right)^2}{64\pi^2}\left(\ln{\left(\frac{m^2(\phi)+\Pi(T,\phi)}{\mu^2}\right)}-\frac{3}{2}\right)-\frac{y^4\phi^4}{16\pi^2}N_fN_{c}\ln{\left(\frac{\phi^2}{\mu^2}\right)}\,.    
 \label{eqn:coleman_wienberg_potential}
\end{equation}
Further, the thermal effective potential can be written as
\begin{align}
  V_{\scaleto{T>0}{5pt}}
  &=
  \frac{T^4}{2\pi^2}\left(J_{\scaleto{\rm B}{4.5pt}}\left(\frac{m^2(\phi)+\Pi(T,\phi)}{T^2}\right)-4N_{f}N_{c}J_{\scaleto{\rm F}{4.5pt}}\left(\frac{y^2\phi^2}{T^2}\right)\right)\,.  
  \label{eqn:thermal_effective_potential}
\end{align}
In the above equations, the mass parameter $m(\phi)$ and the resummed thermal mass $\Pi(T,\phi)$ are defined as:
\begin{align}
m(\phi)^{2}
&=
3\lambda(\mu)\phi^{2}+5\dfrac{\phi^{4}}{\Lambda^{2}},
\qquad
\Pi(T,\phi)
=
\dfrac{T^{2}}{24}\left(6\lambda(\mu)+60\dfrac{\phi^{2}}{\Lambda^{2}}+4y^{2}N^{2}\right)\,.
\end{align}
Finally, the thermal functions $J_{\scaleto{\rm F/B}{6.5pt}}$ used in equation~\eqref{eqn:thermal_effective_potential} are defined in the integral form as follows:
\begin{align}
  J_{\scaleto{\rm F/B}{6.5pt}}(y^2)
  &=
  \int_0^{+\infty}dx\, 
  x^2\ln{\left(1\pm e^{-\sqrt{x^2+y^2}}\right)}\,.
\end{align}
As we are using the RG improved scheme, the quartic $\lambda(\mu)$ runs according to the equation
\begin{equation}
    \lambda(\mu)
    =
    \frac{\lambda_{+}(1-e^b)}
    {1-(\lambda_{+}/\lambda_{-})e^b}\,,
    \quad
    b=\frac{9}{8\pi^2}(\lambda_{+}-\lambda_{-})\ln\frac{\mu}{\mu_0}\,,
\end{equation}
where $\lambda_{\pm}$ are the fixed points for the quartic~(equation~\eqref{eq:quarticfixed}). 

\begin{figure}[t!]
    \centering
    \includegraphics[width=0.75\linewidth]{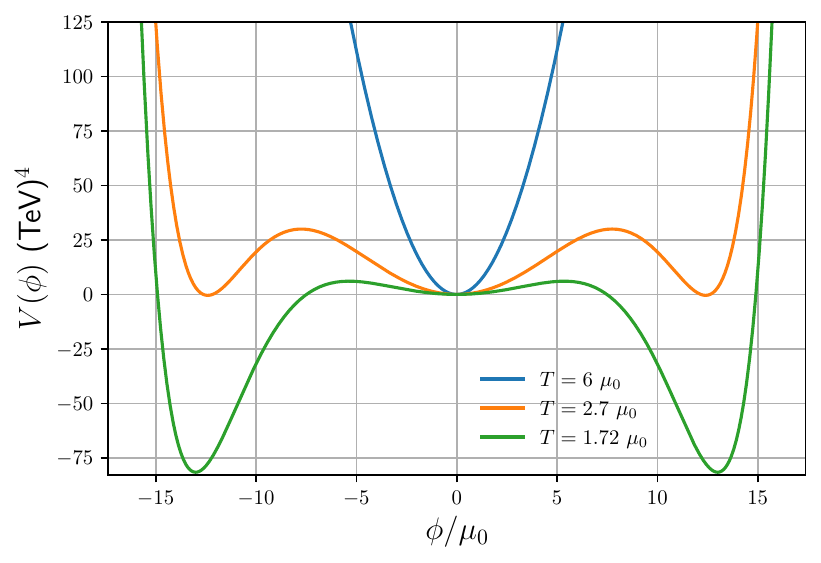}
  \caption{Plot of the potential $V(\phi,T)$ for different values of temperature for the parameter values: $\mu_{0}=500 \text{GeV}$, $\Lambda=10\mu_{0}$, $N=20$ and, $yN=5$. \textcolor{blue}{Blue}: $T=6\mu_{0}$ is a high temperature case with only one vacuum, \textcolor{orange}{Orange}: $T=2.7\mu_{0}$ is the critical temperature, \textcolor{Green}{Green}: $T=1.72\mu_{0}$ is the case with distinct true and false vacua, which is approximately the nucleation temperature for this case.}
    \label{fig:potential}
\end{figure}

The coupling $y$ and the `t Hooft coupling of the gauge theory remain close to their fixed point values. Therefore it is convenient to parametrise our theory by the scale $\mu_0$ at which the quartic coupling $\lambda$ goes through zero. A simple choice of renormalisation scheme is to choose counter-terms in $\overline{\rm MS}$ except for the quartic which is chosen to be:
\begin{equation}
    \delta \lambda = \delta \lambda_{\scaleto{\overline{\rm MS}}{5.5pt}}-\frac{N^2y^4}{16\pi^2}\ln y^2\,.
\end{equation}
This choice results in $(V(\phi,0)/\phi^4)|_{\phi=\mu_0}=0$ up to $\mathcal{O}(\mu_0^4/\Lambda^4)$ corrections. This ensures that $\mu_0$ is the only relevant dimensionful quantity.
For calculation we choose the renormalisation scale $\mu$ such that $\mu=\phi$ if $\phi>T$ and $\mu=T$ otherwise. The functional form of the potential $V(\phi,T)$ is shown in figure~\ref{fig:potential} for different values of the temperature. 
\subsubsection{The Phase Transition: Qualitative Analysis}
Before studying the transition numerically we can make some estimates. 
The thermal correction gives mass to the scalar field $\phi$. The potential including thermal corrections can be roughly written as
\begin{equation}
    V
    \sim \dfrac{1}{2}y^{2}N^{2}T^{2}\phi^{2}+\dfrac{\lambda}{4}\phi^{4}+\dfrac{\phi^{6}}{6\Lambda^{2}}\,.
\end{equation}
This potential has a  minimum at $\phi=0$ and an additional minimum below a certain temperature at
\begin{equation}
    \phi^{2}=\dfrac{\Lambda^{2}}{2}\left(-\lambda + \sqrt{\lambda^{2}-\dfrac{4y^{2}N^{2}T^{2}}{\Lambda^{2}}}\right)\,.
\end{equation}
The critical temperature where the second minimum is degenerate with the $\phi=0$ minimum can be estimated as
\begin{equation}
    T_{c}\sim \dfrac{|\lambda|\Lambda}{yN}\,.
\end{equation}
The vev of the scalar field at $T_c$ is $\langle \phi\rangle \sim \sqrt{|\lambda|}\Lambda$, which implies that the mass of the fermion in the true vacuum is of the order $m_{\psi}\sim y\sqrt{|\lambda|}\Lambda$. Hence,
\begin{equation}
   \dfrac{m_{\psi}}{T_{c}}\sim \sqrt{\dfrac{y^{4}N^{2}}{|\lambda|}}\,.
\end{equation}
The ratio is $m_\psi/T_c\gtrsim 1$ when $|\lambda|\lesssim y^4N^2$. In this regime quarks decouple after the phase transition. We see that for this to be valid the quartic should be of the same order as the quantum corrections from fermion loops and the Coleman-Weinberg mechanism is relevant for the scalar potential. 

The phase transition we study involves a transition in scalar field direction as well as in the gauge theory. Fully studying the transition would involve understanding the gauge theory in the strongly coupled regime. Lattice gauge theory results are not directly applicable to our case due to the presence of fermions with $\phi$-dependent mass. However, the impact of the gauge sector on our transition is expected to be subdominant to the scalar sector, essentially due to the fact that the gauge sector does not have supercooling by itself. 

A simple estimate to see this is to compare the $O(3)$ symmetric bounce actions of both transitions close to their critical temperature. The scalar quartic coupling and the Yukawa coupling follow large-$N$ counting, so we can compare the scalar field bounce action with the gauge theory action using large-$N$ arguments. Near $T_c$, the action can be estimated in the thin-wall limit and is given by
\begin{align}
    \frac{S_3}{T}\sim \frac{16\pi}{3}\frac{N^2}{(yN)^3}\frac{1}{\epsilon (T)^2}\,.
\end{align}
Comparing with equation~\eqref{eqn:o_3_ym} we see that the scalar contribution strongly dominates the action for $yN$ within  perturbative control. 

\subsubsection{The Phase Transition: Full Analysis} 
\label{sec:numerical_analysis}
\begin{figure}[t!]
    \centering
    \begin{minipage}{0.48\linewidth}
            \includegraphics[width=\linewidth]{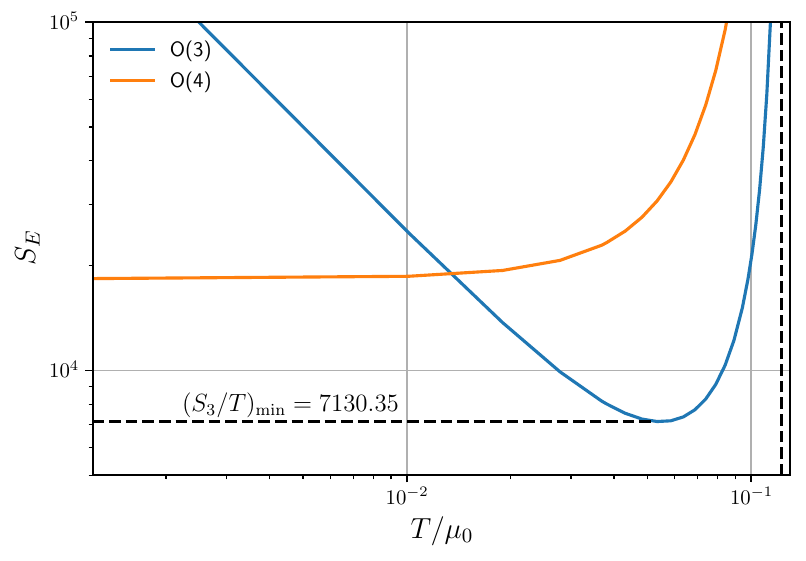}
    \end{minipage}    
    \begin{minipage}{0.48\linewidth}
        \includegraphics[width=\linewidth]{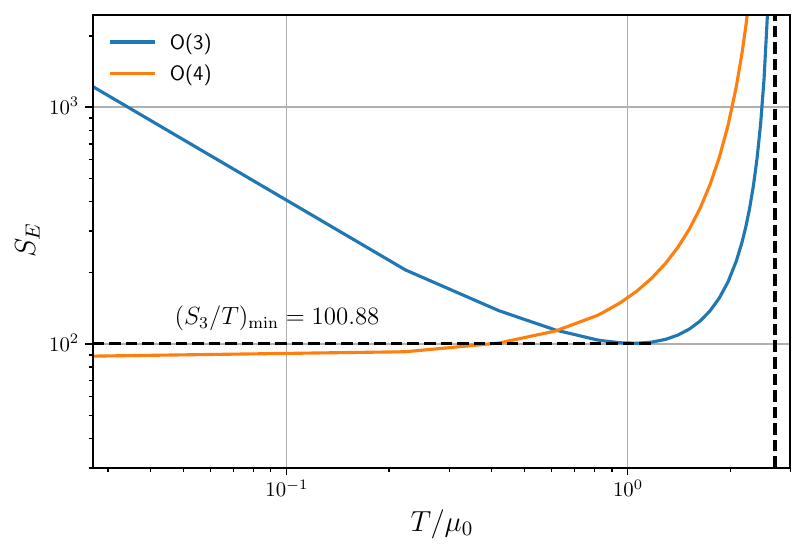}
    \end{minipage}
    \caption{The $O(3)$ and $O(4)$ symmetric bounce actions evaluated for $\Lambda=5$ TeV, $\mu_0=500$ GeV, $N=20$ and \textbf{Left}: $yN=4$ where and $T_{c}=0.12 \mu_{0}$. In this case, the transition does not complete. \textbf{Right}: $yN=5$ where $T_{c}=2.71 \mu_{0}$. The vertical dashed line represents the critical temperature $T_{c}$ where the action $S_{\scaleto{\rm E}{4pt}}$ asymptotes to infinity.}
    \label{fig:actions}
\end{figure}

To study the phase transition we evaluate the bounce action numerically using the Python module {\tt CosmoTransitions} \cite{Wainwright:2011kj}. The results are given in figure~\ref{fig:actions}. The qualitative behavior of the Euclidean action is as follows: Around the critical temperature $T\sim T_c$ both $O(4)$ and $O(3)$ symmetric bounce actions are very large due to the small difference between false and true vacuum energies, with the $O(3)$ symmetric bounce dominating the transition. At very low temperature $T\ll T_c$ the $O(4)$ symmetric action becomes constant in $T$   approaching its zero temperature value while the $O(3)$ symmetric action increases approximately as $1/T$.
Hence, the $O(4)$ bounce dominates in this regime. The $O(3)$ symmetric action has a minimum that is located a factor of few away from $T_c$. Therefore, if the transition happens and is dominated by $O(3)$ bounce, the minimal value of the ratio $T_n/T_c$ is controlled by the location of the minimum, in the parameter space of our interest this is given by $T_n/T_c\gtrsim \mathcal{O}(0.1)$.\\
In the left panel of figure~\ref{fig:actions}, we have shown the case where the parameters take the value: $\Lambda=5$ TeV, $\mu_{0}=500$ GeV, $yN=4$ and $N=20$. In this scenario, as the plots show, the minimum value of the Euclidean action is around $S_{\scaleto{\rm E}{4pt}}\approx 7130$. This is an order of magnitude larger than 120, the value required for the bubbles to nucleate and the transition to move forward at around the electroweak scale. Hence, in this case, the nucleation condition is not satisfied. As the phase transition cannot complete, the universe is trapped in an eternally inflating CFT-like phase, which corresponds to a YM theory with massless quarks coupled to a single scalar.  

Next, we consider parameters for which phase transition can complete. This is shown in the right panel of figure~\ref{fig:actions} where the parameters for the theory take the values: $\Lambda=5$ TeV, $\mu_0=500$ GeV, $yN=5$, and $N=20$. One can see that the minimum of the Euclidean action is $S_{\scaleto{\rm E}{4pt}}\approx 100$ and hence the transition can complete. As the phase transition completes, the next thing to check is the $m_{\psi}/T_{n}$ ratio to see if the fermions are decoupled inside the nucleated bubble.

\begin{figure}[t!]
    \centering
    \includegraphics[width=0.75\linewidth]{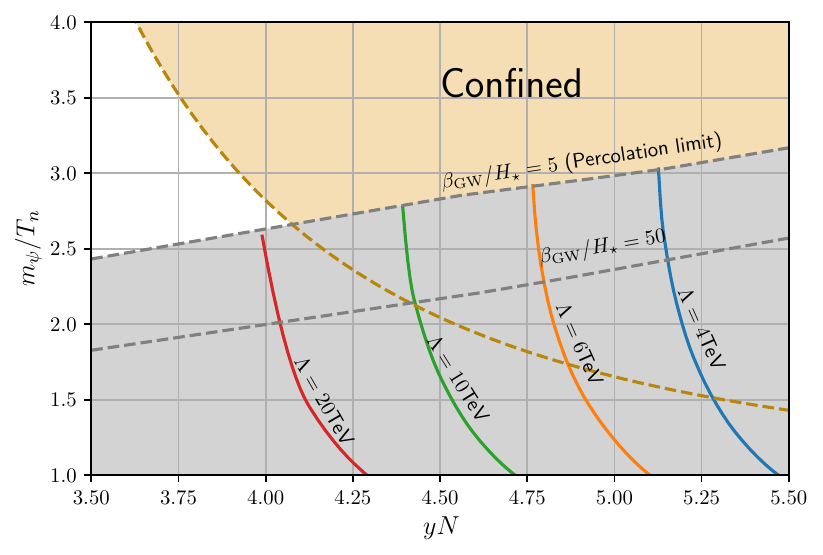}
    \caption{The ratio of the fermion mass after the transition to nucleation temperature $m_\psi/T_n$ as a function of $yN$. The region above the golden dashed line represents the parameter space in which after transition YM theory is confined corresponding to $g^2N_c\to~\infty$. The grey patch represents the parameter space in which percolation happens where we have plotted contours corresponding to $\Lambda=20,~10,~6$ and 4 TeV. As one move towards the left in plot above, one finds to be in the region where the `t Hooft coupling is small and hence we do not have confinement inside the bubble during the scalar field transition. The figure is obtained for $\mu_0=500$ GeV, $N=20$.}
    \label{fig:mpsivsTn}
\end{figure}

Figure~\ref{fig:mpsivsTn} shows ratio of the fermion mass after transition to nucleation temperature $m_\psi/T_n$ as a function of $yN$ for $N=20$,  $\mu_0=500$ GeV. The grey shaded patch represents the parameter space accessible by our model in which percolation happens. In this region, we have plotted the contours for some fixed values of $\Lambda$ as shown in the figure. The shaded region above the golden dashed line represent the parameter space in which inside the phase transition bubble the YM theory is confined, where we define the confinement scale as $g N_{c}^{2}\rightarrow \infty$. The condition is applied using the three-loop beta function of YM theory\cite{ParticleDataGroup:2024cfk}. Since perturbative control is lost close to the confinement scale, detailed quantitative conclusions cannot be drawn directly from the figure, especially near the boundary of the regions. However, qualitatively, far from boundaries the precise definition of confinement will not be relevant and the theory will confine immediately after the phase transition in the interior of the shaded region, and will remain deconfined just after the transition in the unshaded region.

\begin{figure}[t!]
\begin{minipage}[t]{0.47\linewidth}
      \centering
    \includegraphics[width=\linewidth]{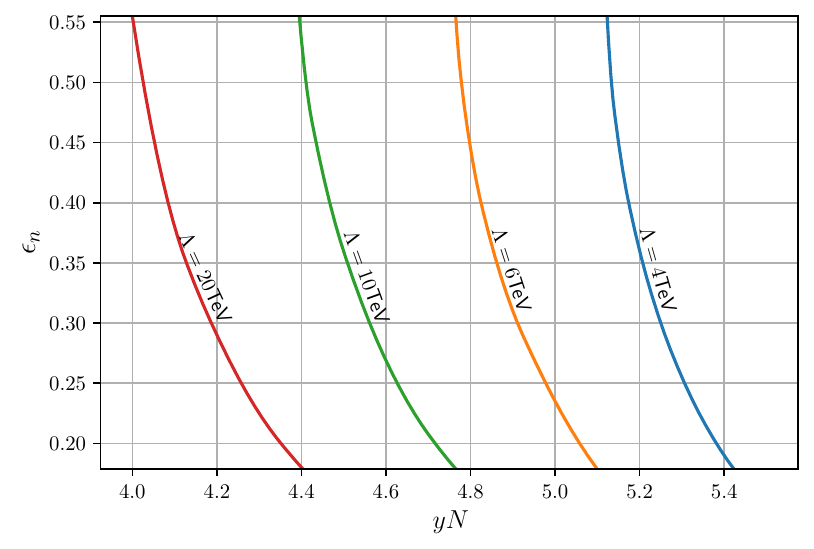}
    \caption{Supercooling parameter $\epsilon_n$ as a function of $yN$ for different values of $\Lambda$, fixing $\mu_{0}=500$GeV and $N=20$.}
     \label{fig:epsilon_vs_yN}
\end{minipage}
\hspace*{0.25cm}
\begin{minipage}[t]{0.47\linewidth}
      \centering
    \includegraphics[width=\linewidth]{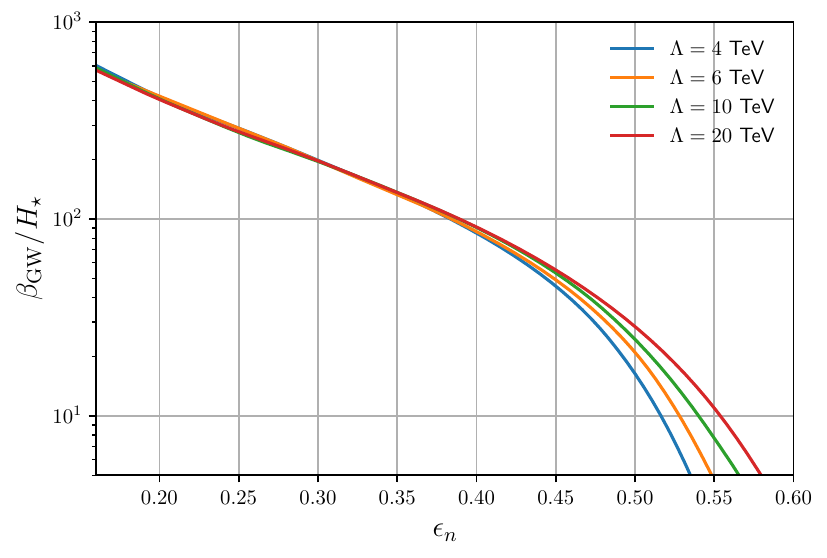}
    \caption{Inverse duration of the transition normalized with the Hubble constant $\betagw/H_{\star}$ as a function of  $\epsilon_n$ for different values of $\Lambda$, fixing $\mu_{0}=500$GeV and $N=20$.}
     \label{fig:beta_vs_epsilon}
\end{minipage}
\end{figure}

We now study whether in these regions of parameter space the phase transition is supercooled or not. We plot the supercooling parameter $\epsilon_{n}$ as defined in equation~\eqref{eq:epsilon_n} as a function of $yN$ for different values of $\Lambda$ in figure~\ref{fig:epsilon_vs_yN}. The figure shows that for every value of $\Lambda$ (corresponding to the contours plotted in figure~\ref{fig:mpsivsTn}) there exists a choice of $yN$ where the supercooling parameter $\epsilon_{n}\sim \mathcal{O}(1)$. Further, as discussed in section~\ref{sec:appendixGW} the important parameter that sets the strength of GW emission is $\betagw$. Figure~\ref{fig:beta_vs_epsilon} shows the value of $\betagw$ calculated using equation~\eqref{eq:beta} plotted against the value of $\epsilon_{n}$. The plot shows a strong inverse correlation between $\epsilon_{n}$ and $\betagw$. We also studied the percolation criteria given in section~\ref{sec:appendixGW} and show that in our parameter space, percolation happens for $\betagw /H_{\star}> 5$. 

The numerical analysis presented above show us that there is a parameter space in our model where the phase transition completes, the fermions become massive enough to decouple inside the nucleated bubble leading to confinement, and this can happen with large supercooling ($\epsilon_{n}\sim \mathcal{O}(1)$ and $\betagw/H_{\star} \sim O(10)$). There can be two other qualitatively different cases after the phase transition. In the first of these two $m_{\psi}/T_{n}\lesssim 1$. In this case, the phase inside the bubble has massive fermions but the theory is deconfined. As $m_{\psi}$ is close to $T_{n}$ the $\beta$-function of the gauge coupling runs very fast and soon after scalar transition ends the theory confines. In this scenario, the two phase transitions are close and not well separated. The second possible scenario is when $m_{\psi}/T_{n}\ll 1$. In this case the confinement transition and the scalar transition are well separated. These three cases are summarised in figure~\ref{Cartoon_three_bubbles}. 

\section{Gravitational Waves in Our Model}
\label{sec:causal_tails}
\begin{figure}[t]
    \centering
    \begin{subfigure}[t]{0.3\linewidth}
    \centering
      \includegraphics[width=\linewidth]{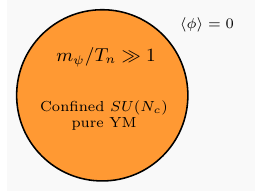}  
      \caption{Case 1}
       \label{fig:bubble1}
    \end{subfigure} 
    \begin{subfigure}[t]{0.3\linewidth}
    \centering
      \includegraphics[width=\linewidth]{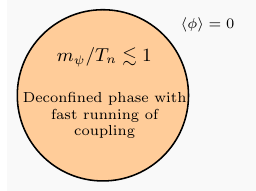}
      \caption{Case 2}
         \label{fig:bubble2}
    \end{subfigure} 
    \begin{subfigure}[t]{0.3\linewidth}
    \centering
      \includegraphics[width=\linewidth]{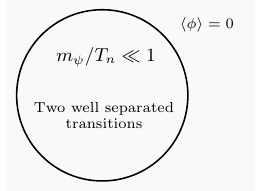} 
      \caption{Case 3}
         \label{fig:bubble3}
    \end{subfigure} 
    \caption{Three different scenarios depending on the ratio $m_{\psi}/T_{n}$ at the nucleation temperature. In the region outside the bubble, $\langle \phi\rangle=0$ and the fermions are massless. Inside, there are three possibilities: \textbf{1.~$m_{\psi}/T_{n}\gg 1$}:  The true vacuum inside is a  pure $SU(N_{c})$ YM theory in the confined phase. \textbf{2.~$m_{\psi}/T_{n}\lesssim 1$}: The true vacuum is deconfined with massive fermions, with a large $\beta$-function. In this case, the scalar phase transition is closely followed by the confining transition. \textbf{3.~ $m_{\psi}/T_{n}\ll 1$}: The true vacuum remains close to the fixed point until the universe cools substantially down to $T\sim m_\psi$.}
    \label{Cartoon_three_bubbles}
\end{figure}

In the previous section, we demonstrated that for some parameter space in our model we obtain a supercooled confinement phase transition (corresponding to figure~\ref{fig:bubble1}). At  nucleation, the temperature inside the bubble in this case is below the confinement scale and triggers a fast confining transition. The bubble wall separating the true and false vacua consists of the scalar as well as the gauge theory domain wall. Bubble collisions are dominated by the scalar with sub-leading effects due to the gauge theory component of the walls. The sound waves and turbulence are generated in the true vacuum plasma, which may be in the confined phase or a strongly-coupled phase close to confinement.  We see that in this case the phase transition has the similar GW signature as that of first order phase transitions driven by scalar fields. We leave the detailed study of these sub-leading effects for future work. As noted in section~\ref{sec:appendixGW}, an important parameter characterizing the strength of the GW signal is the bubble wall velocity $v_{w}$. In case of strongly supercooled phase transitions, it is well-known that the bubble walls achieve relativistic velocities\cite{Espinosa:2010hh}. We checked that the runaway wall condition is satisfied for strongly supercooled transitions in our model~\cite{Bodeker:2009qy}. The exact determination of wall velocity will require modeling the properties of strongly-coupled plasma and will depend on the model e.g. the decay modes of confined particles. (see \cite{Baldes:2020kam,Bachmaier:2023wzz,Gouttenoire:2021kjv,Baldes:2021aph} regarding dynamics of confining domain walls) 

The second possibility, as shown in figure~\ref{fig:bubble2}, is when the theory remains in the strongly coupled deconfined phase inside the bubble. However the fermions decouple and the gauge coupling runs very fast with interesting observable consequences. Due to the large $\beta$ function, the equation of state parameter $w$ deviates significantly from $1/3$ and has a time dependence which may be observable in the causal tails of the GW spectrum.

\begin{figure}[t]
    \centering
\includegraphics[width=0.75\linewidth]{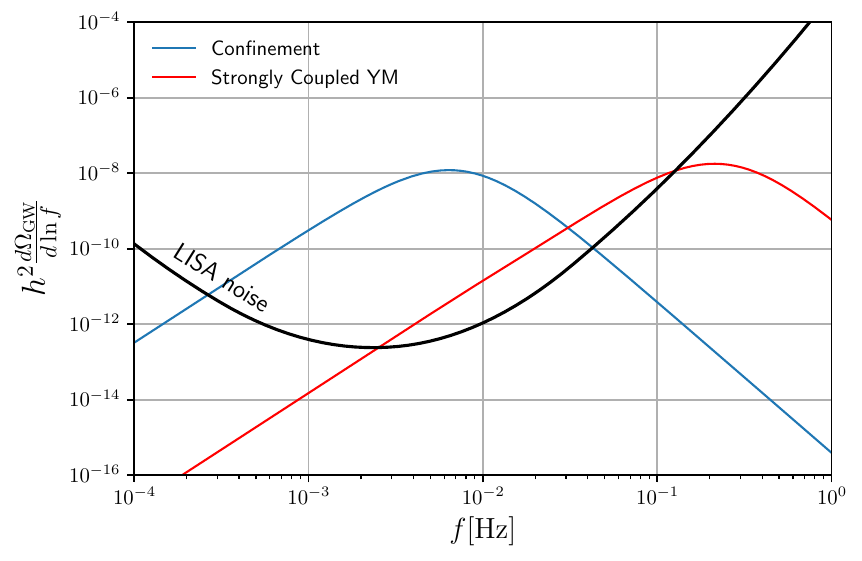}
    \caption{The GW spectra for two benchmarks. {\bf Blue}: The benchmark parameters \{$yN=4.8$, $N=20$, $\mu_{0}=500$ GeV, $\Lambda=6$ TeV\}; the transition temperature is $T_{\star}=600$ GeV with $\alpha=2.3$, $\betagw/H_{\star}=50$. This corresponds to case 1 in figure~\ref{Cartoon_three_bubbles}.
    {\bf Red}: The benchmark parameters \{$yN=3.3$, $N=20$, $\Lambda=300\mu_0$, $\mu_{0}=5$  TeV\}; the transition temperature is  $T_{\star}=10^{5}$ GeV with $\alpha=1$, $\betagw/H_{\star}=10$. This corresponds to case 2 in figure~\ref{Cartoon_three_bubbles}.  The LISA noise curve is adapted from~\cite{Babak:2021mhe}. 
    }
    \label{fig:benchmarkpoints}
\end{figure}
Under standard assumptions, the GW spectrum sourced by a phase transition roughly peaks around $f \sim \betagw/v_{w}$ falling as $1/f$ for high frequencies and $f^3$ for low frequencies~\cite{Caprini:2009yp}. 
The peak and the high-frequency behaviour depend on the details of the phase transition while the low-frequency tail is fixed by causality making it a model-independent probe of cosmology at that epoch \cite{Cai:2019cdl,Hook:2020phx}.

For the low-frequency modes, one can further separate the two cases as sub-horizon modes where $f\gg \mathcal{H}_{*}$ and the super-horizon modes $f\ll \mathcal{H}_{*}$, where $\mathcal{H}_{*}$ is the conformal Hubble scale when the phase transition completes. The superhorizon modes are the ones which carry imprint of the background cosmology and deviate from the $f^3$ scaling. There are two competing effects that affect the spectrum. These modes are over-damped and hence their production is suppressed. On the other hand, since they are super-horizon, they do not red-shift until they enter the horizon which increases the power stored in them relative to e.g.~background radiation.  Due to this, as shown in \cite{Hook:2020phx}, the spectrum of observed GWs would have a distinct feature at the frequency corresponding to $\mathcal{H}_*$. We briefly review the causal tails of GWs and their dependence on the equation of state parameter $w$ \cite{Hook:2020phx, Brzeminski:2022haa} in appendix~\ref{appendix:causal_tail_review}. For a constant equation of state parameter $w$, the modified GW spectrum takes the form:
\begin{equation}
    \dfrac{d\Omega_{\rm GW}}{d\ln f}\sim f^{3-2\left(\frac{1-3w}{1+3w}\right)}\,,~~~(f\ll \mathcal{H}_{*})\,.
\end{equation}

Figure~\ref{fig:benchmarkpoints} represents the GW spectrum for two benchmark points. The spectrum is obtained assuming that the GWs are dominantly produced by sound waves. The first benchmark point represents the phase transition with the confined YM theory in the true vacuum. The second benchmark point is chosen in such a way that after the scalar field transition, we are left with a strongly coupled YM theory which runs and confines at lower (but comparable to scalar phase transition) temperature. 
The benchmark was chosen to have a higher temperature phase transitions so that a longer part of tail is present above the LISA noise curve, amenable to studying the modification of the causal tail of the spectrum.

\subsection{Causal Tails in Our Model}
\begin{figure}[t]
    \centering
    \includegraphics[width=0.75\linewidth]
    {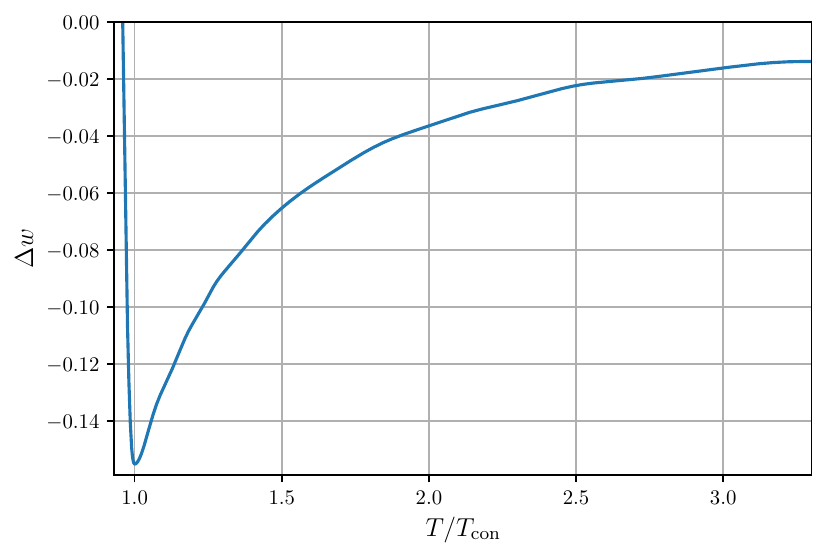}
    \caption{The deviation of the equation of state parameter $w$ for a pure YM theory from $w = 1/3$ as a function of $T/T_{\rm con}$, where $T_{\rm con}$ is the confinement temperature of the YM theory~\cite{Panero:2009tv}. We have assumed $g_{*,\scaleto{\rm SM}{5pt}} = 106$, and the number of colors in the YM theory, $N_c = 10$.} 
    \label{fig:lattice_w}
\end{figure}

We now consider a theory with a strong coupling and its impact on the causal tail spectrum. For this, we first take a qualitative look at the lattice results that indicate the temperature (or time) dependence of the equation of state parameter $w$. This dependence is explicitly shown in figure~\ref{fig:lattice_w} (see \cite{Panero:2009tv} for details). At very large $T/T_{\rm con}$, where $T_{\rm con}$ is the temperature at which the theory confines, the parameter $\Delta w$ goes to zero with respect to $w=1/3$ showing that the system behaves like radiation. As $T/T_{\rm con}$ approaches one, $(-\Delta w)$  grows fast and abruptly drops below the confinement temperature. 
However, the behaviour below $T_{\rm con}$ depends on additional model building details such as the interaction between the dark and the visible sectors. We are ultimately interested in having scenarios in which the sectors are in thermal equilibrium. For this reason, we assume that the same interaction between the two sectors will allow the decay of glueballs.

\begin{figure}[t]
\centering
\begin{subfigure}{0.75\linewidth}
    \centering
    \includegraphics[width=\linewidth]{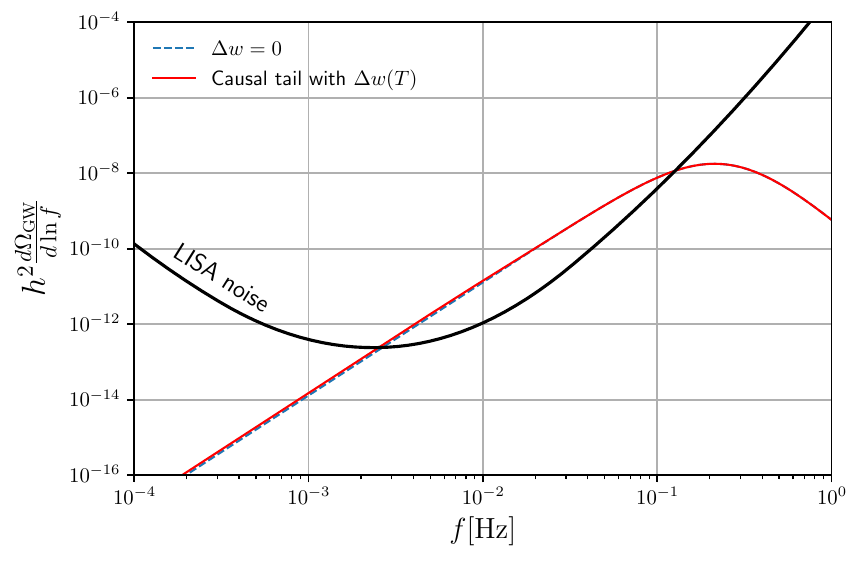}  
\end{subfigure}
\\
\begin{subfigure}{0.75\linewidth}
    \centering
    {
    \hspace{-7mm}
    \includegraphics[width=\linewidth]{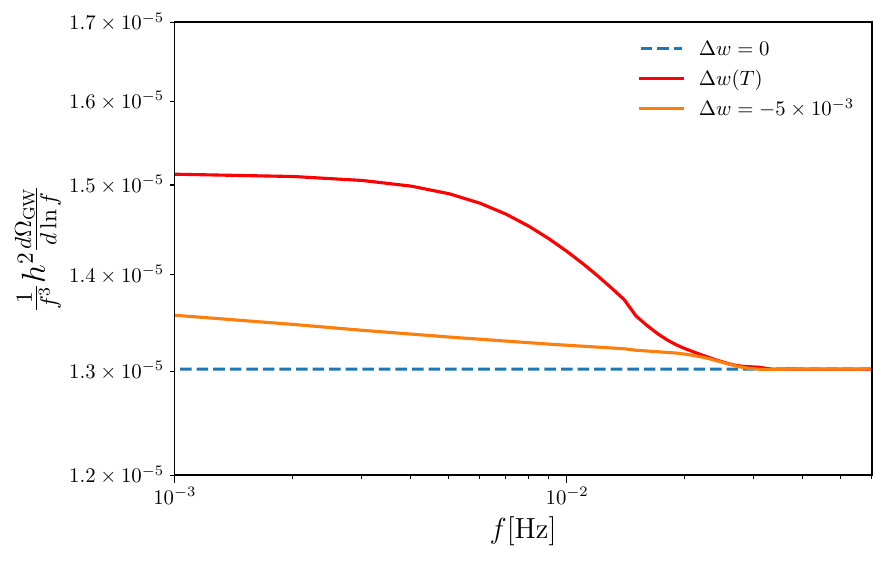} 
    }
\end{subfigure}
    \caption{Modification of causal tails for $w(T)$ in figure~\ref{fig:lattice_w}. \textbf{Left}: The GW spectrum from sound waves (red) for the benchmark \{$yN=3.3$, $N=20$, $\Lambda=300\mu_0$, $\mu_{0}=5$ TeV\}, which has $\betagw/H_{\star}=10$,  $\alpha=1$ and $v_{w}\sim 0.9$, $T_{\star}=10^{5}$~GeV followed by a confinement transition at $T_{\rm con}=3\times 10^4$~GeV. The blue dashed line corresponds to $w=1/3$. \textbf{Right}: Deviation of the spectrum (red) from $f^3$ scaling (blue dashed). We include a reference curve (orange) corresponding to $\Delta w=-5\times 10^{-3}$. }
    \label{fig:causal_tails}
\end{figure}

We now use this time-dependent $w$ to solve the Friedmann equations and equation of motion of the metric fluctuations given in equation~\eqref{eqn:graviton_eom_approx_1} numerically and study the change in the GW spectrum. We assume the sensitivity estimates reported in \cite{Hook:2020phx}. For simplicity we assume that just right after the confinement the energy density of the universe is radiation dominated. This ignores the departure from radiation dominant era during the confinement phase transition  as well as the effect of glueballs after confinement. These effects would plausibly make the signal stronger and are left for future work.

In figure~\ref{fig:causal_tails} we show the modification of the GW spectrum due to temperature dependent $w(T)$. The blue dashed line is the unmodified spectrum produced by sound waves taken from equation~\eqref{eqn:gw_fraction_sound_waves} with $\betagw/H_{\star}=10$, $v_{w}\sim 1$ and $\alpha=1$. Further, this plot is corresponding to phase transition at a temperature $T_{\star}=10^{5}$ GeV. The black curve shows the LISA sensitivity that has been plotted according to the data in \cite{Babak:2021mhe}. The red curve shows the resulting GW spectrum considering temperature dependent $\Delta w(T)$ as plotted in figure \ref{fig:lattice_w}. In the right panel, we show the departure of the GW spectrum from the $f^{3}$ scaling of causal tails.  The red curve initially deviates from $f^{3}$ scaling and then becomes parallel to $d\Omega/d\ln f\sim f^3$ line. The orange line corresponds to a case with $\Delta w=-5\times 10^{-3}$ which, according to \cite{Brzeminski:2022haa} is detectable at LISA for these values of temperature and $\betagw$. As this is significantly smaller modification compared to the case corresponding to temperature dependent $w(T)$, we see that the deviation in the red curve is larger than the estimated sensitivity. For a loud enough GW signal, we may hope to extract properties of the strongly-coupled phase in this case.

\section{Conclusion}
\label{sec:discussion}

In this paper we studied the possibility of obtaining a supercooled confinement phase transition able to generate a strong GW signal. After reviewing different approaches and attempts to obtain strong supercooling we note that surprisingly, even though confinement transitions can be strongly first order (e.g.~in large-$N$ gauge theories), they are not supercooled, and consequently do not produce a strong GW signal.

The RS model provides a calculable framework for strongly coupled gauge theories where supercooling is possible. However, the precise 4D gauge theory that is dual to the RS model is not known, and in the light of the difficulty in obtaining supercooled confinement in the controlled examples in 4D above motivates looking for a 4D analog that adapts the features of the RS model that allow supercooling.

We constructed an explicit model that builds these features and achieves supercooling in a confining transition. The gauge theory is a strongly coupled theory close to an IR fixed point. This is coupled to a scalar that undergoes a thermal phase transition and triggers confinement. In a part of the parameter space, the confinement and scalar transition occur at the same time. It is also generically possible to have the transitions be separated in the cosmic history. Interestingly, when their separation is not large, their GW spectrum has unique features within the reach of future observatories.

It is worth noting the similarity of the scalar field transition to a chiral symmetry breaking transition. The scalar field acts as a simple linear sigma model for the chiral symmetry breaking. Indeed, models where the scalar does not arise as an external field but is a composite of the gauge theory itself are probably the most compelling versions of this framework. It will be interesting to pursue more explicit constructions of such models. Models with walking technicolor~\cite{Miura:2018dsy, Azatov:2020nbe} can be related constructions in this direction.

We can extract some qualitative lessons from our example. In our model, the transition is supercooled for a specific narrow range of a `t Hooft coupling at the fixed point. In a strongly coupled theory with $N_c$ colors and $N_f$ flavors, we thus want to be inside the conformal window, with the ratio $N_f/ N_c$ tuned to give us the appropriate `t Hooft coupling at the fixed point. Therefore, we can very roughly imagine that for $N_f /N_c$ much smaller than this value the phase transition qualitatively resembles the transition in pure gauge theory and is not supercooled. For a larger ratio the transition does not complete. In general, it will be very interesting to study the phase diagram of gauge theory not only in terms of the equilibrium phases but also the metastable phases. We see that in the present context, these phases drive the phenomenology of GWs.

Even though there is a large body of work on GWs from early universe phase transitions, there are many open questions that deserve a detailed study.  In the case of a strongly-coupled phase transition, even the basic nature of the phase transition itself is only well-understood for isolated cases. The evolution of the gauge theory domain walls, the dynamics of the confining plasma, e.g.~including glueballs and other hadronic states, effect of additional chiral matter, and the $\theta$-angle are all open questions. These aspects will need to be better understood to have robust quantitative predictions for GWs from phase transitions.

{\bf Note Added:} While finishing the write up of this paper we became aware of the reference~\cite{Fujikura:2025iam} which overlaps with some of the ideas presented in this paper.
\section*{Acknowledgement}
We thank Ed Hardy, Saquib Hassan, John March-Russell, David Mateos, Fabrizio Rompineve, Mike Teper, Kieran Twaites for useful discussions. We also thank Georges Obied for illuminating discussions about boiling water and $\mathcal{N}=4$ SYM. GRK would like to express gratitude towards Somerville College, Oxford, and The Clarendon Fund, Oxford for supporting the research via the Oxford Ryniker Lloyd Graduate Scholarship jointly with the Clarendon Fund Scholarship. The work of PA and VL is supported by the STFC grant ST/X000761/1. For the purpose of Open Access, the author has applied a CC BY public copyright licence to any Author Accepted Manuscript version arising from this submission.

\appendix

\section{Causal Tails}
\label{appendix:causal_tail_review}
Here we briefly review the results of \cite{Cai:2019cdl,Hook:2020phx,Brzeminski:2022haa}. 
The equation of motion for the graviton modes in comoving coordinates for a comoving mode $k$ is given by\cite{Caprini:2009yp}:
\begin{equation}
    \dfrac{\partial^{2} h_{ij}}{\partial \tau^{2}}+2\mathcal{H}\dfrac{\partial h_{ij}}{\partial \tau}+k^{2}=a^{2}\dfrac{32\pi G\rho}{3}\Pi_{ij}=J_{ij}~~,
    \label{eqn:graviton_eom}
\end{equation}
where $\Pi_{ij}$ is the anisotropic stress tensor. For the two independent polarizations $(+,\times)$, this can be simplified in terms of the graviton and source amplitude $h$ and $J$ respectively. If the phase transition happens at conformal time $\tau_{*}$, then one considers $h(\tau)=0$ for $\tau<\tau_{*}$, i.e. there is no production of waves before the completion of phase transition. Further, from the perspective of the small $k$ modes where $k\ll \betagw/v_{w}$, one can assume that phase transition happens very fast and hence the source can be approximated as: $J(k,\tau)\propto \delta(\tau-\tau_{*})$. In addition to this, for these modes, the source is very localized, and hence the spatial Fourier transform can be approximated to be a constant $J_{*}$. In this case, the modes are solution to the differential equation\cite{Hook:2020phx}:
\begin{equation}
\partial_{\tau}^{2}h+2\mathcal{H}\partial_{\tau}h+k^{2}h=0,~~h(k,\tau_{*})=0,~~\partial_{\tau}h(k,\tau_{*})=J_{*}~~.
\label{eqn:graviton_eom_approx_1}
\end{equation}
The Hubble $\mathcal{H}(\tau_{*})\equiv \mathcal{H}_{*}$ is considered to be constant in \cite{Hook:2020phx} during the period of GW production. Further, while analyzing the above for super-horizon modes with $k\ll \mathcal{H}_{*}$ one can further approximate the equation as follows: in the small time interval (say one e-fold) starting from $\tau_{*}$, in the linear limit, one can write $h(k,\tau_{*}+\mathcal{H}_{*}^{-1})\approx J_{*}/\mathcal{H}_{*}$. Along with this, these modes when being produced will be over-damped and hence will not oscillate giving us $\partial_{\tau}h(k,\tau_{*}+\mathcal{H}_{*}^{-1})=0$. As these modes have to wait for much longer times to enter the horizon, the initial conditions in equation~\eqref{eqn:graviton_eom_approx_1} can be modified as:
\begin{equation}
 h_{\rm super}(k,\tau_{*})\sim \dfrac{J_{*}}{\mathcal{H}_{*}},~~\partial_{\tau}h_{\rm super}(k,\tau_{*})\approx 0~~.
\end{equation}
Once these modes enter the horizon, a crude approximation of turning off the friction term in equation~\eqref{eqn:graviton_eom_approx_1} then gives:
\begin{equation}
    h_{\rm super}(k,\tau)\sim \dfrac{a(\tau_{k})}{a(\tau)}\dfrac{J_{*}}{\mathcal{H}_{*}}\sin k\tau \,,
\end{equation}
where $\tau_{k}$ is the conformal time when $k=\mathcal{H}(\tau_{k})$. As presented in the \cite{Hook:2020phx}, solving the equation~\eqref{eqn:graviton_eom_approx_1} for a general equation of state $p=w \rho$ where the scaling constant and the conformal Hubble take the form:
\begin{equation}
    a(\tau)\propto \tau^{n},~~\mathcal{H}=\dfrac{n}{\tau},~~n=\dfrac{2}{1+3w}~~,
\end{equation}
the super-horizon mode amplitude scales like:
\begin{equation}
    h_{\rm super}\approx \dfrac{J_{*}\tau_{*}}{(k\tau)^{n}}\sin k\tau. 
    \label{eqn:super_horizon_amplitude_scaling}
\end{equation}
The spectrum for GWs $d\Omega_{\scaleto{\rm GW}{4pt}}/d\log k$ is defined as:
\begin{equation}
    \dfrac{d\Omega_{\scaleto{\rm GW}{4pt}}}{d\log k}=\dfrac{1}{\rho_{c}}\dfrac{d\rho_{\scaleto{\rm GW}{4pt}}(k,\tau)}{d\log k}=\dfrac{1}{\rho_{c}}\dfrac{k^{5}P_{h}(k,\tau)}{2(2\pi)^{3}a^{2}G}\,,
    \label{eqn:gw_spectrum_general}
\end{equation}
where $\rho_{c}$ is the critical energy density and $P_{h}$ is the power spectrum defined as:
\begin{equation}
    \langle h(\mathbf{k},\tau)~h(\mathbf{k}',\tau)\rangle=(2\pi)^{3}\delta^{(3)}(\mathbf{k}-\mathbf{k}')P_{h}(k,\tau)~~.
\end{equation}
The factors of $k$ in equation~\eqref{eqn:gw_spectrum_general} can be counted as follows: (a) two from the derivatives in the definition of $\rho_{\scaleto{\rm GW}{4pt}}(\mathbf{x},\tau)$ which then give the factor of $k^{2}$ when expressed in terms of the Fourier transform $\rho_{\scaleto{\rm GW}{4pt}}(k,\tau)$, (b) one from the derivative with respect to $\log k$, and (c) two from the phase space factor giving the overall multiplicative factor of $k^{5}$. For a general equation of state, using the $k$ scaling of $h_{\rm super}$ as in equation~\eqref{eqn:super_horizon_amplitude_scaling}, and substituting it in equation~\eqref{eqn:gw_spectrum_general} one gets the scaling:
\begin{equation}
    \dfrac{d\Omega_{\scaleto{\rm GW}{4pt}}}{d\log k}\sim k^{3-2\left(\frac{1-3w}{1+3w}\right)}\,,~~~(k\ll \mathcal{H}_{*})\,.
\end{equation}
When $w=1/3$, i.e., a radiation dominated epoch, one retrieves the $k^{3}$ scaling. On the other hand, when $w\neq 1/3$, the scaling of the super-horizon modes is different from the $k^{3}$ scaling and hence gives a distinctive feature to the tail of the spectrum that allows us to understand the Hubble value during the phase transition in a model independent fashion (as we don't consider any specific details or the dynamics of the source creating the GW waves) and hence can help us understand the physical processes at the epoch.

\bibliographystyle{utphys}
\bibliography{references}
\end{document}